# **Bitcoin MiCA White Paper**[1]

Juan Ignacio Ibañez[a]

Lena Klaaßen[b]

Ulrich Gallersdörfer[b]

Christian Stoll[b]

*V1.0: 14 March 2024*

---

[a] UCL Centre for Blockchain Technologies, http://blockchain.ucl.ac.uk/, author's email: j.ibanez@ucl.ac.uk

[b] CCRI - Crypto Carbon Ratings Institute, https://carbon-ratings.com, email: hi@carbon-ratings.com

# Bitcoin MiCA White Paper

**Disclaimer:** This document is written as an academic exercise, with the goal of exploring the feasibility of writing a white paper in accordance with Regulation (EU) 2023/1114 (MiCA). It is meant as a Proof of Concept (PoC) illustrating a concrete application of the requirements of MiCA. Like the MiCA white papers PoC shared by ESMA[2], this document is solely for the purposes of the PoC, to inform the public as to how a crypto-asset white paper could work, inspire public debate and feedback, and enhance the public conversation around the implementation of EU regulations. The authors of this white paper do not accept liability or responsibility in relation to the published material as the purpose of the PoC is to showcase how a white paper could be written, taking the world's most important crypto-asset project as an example.

We, the authors of this white paper, are neither offerors, developers, advisors, or service providers of this crypto-asset, nor can be considered a part of this crypto-asset project. This white paper is written solely for educational, academic and scientific purposes.

The authors of this white paper are not an offeror, persons seeking admission to trading, nor operators of the trading platform. Nevertheless, we declare that, to the best of our knowledge, the information presented in the white paper is fair, clear and not misleading, and the whitepaper makes no omission likely to affect its import.

---

[2] https://www.esma.europa.eu/document/mica-white-papers-poc

# Bitcoin MiCA White Paper

2024-03-14
|[N: I.01]

**N: 1.00 Table of contents**

## I. Compliance with duties of information

| N | Field | Content |
|---|---|---|
| I.02 | **Statement in accordance with Article 6(3) of Regulation (EU) 2023/1114** | This crypto-asset white paper has not been approved by any competent authority in any Member State of the European Union. The offeror of the crypto-asset is solely responsible for the content of this crypto-asset white paper. |
| I.03 | **Compliance statement in accordance with Article 6(6) of Regulation (EU) 2023/1114** | This crypto-asset white paper complies with Title II of Regulation (EU) 2023/1114 and, to the best of the knowledge of the management body, the information presented in the crypto-asset white paper is fair, clear and not misleading and the crypto-asset white paper makes no omission likely to affect its import. |
| I.04 | **Statement in accordance with Article 6(5), points (a), (b), (c) of Regulation (EU) 2023/1114** | The crypto-asset may lose its value in part or in full, may not always be transferable and may not be liquid. |
| I.05 | **Statement in accordance with Article 6(5), point (d) of Regulation (EU) 2023/1114** | false |
| I.06 | **Statement in accordance with Article 6(5), points (e) and (f) of Regulation (EU) 2023/1114** | The crypto-asset is not covered by the investor compensation schemes under Directive 97/9/EC of the European Parliament and of the Council.<br><br>The crypto-asset is not covered by the deposit guarantee schemes under Directive 2014/49/EU of the European Parliament and of the Council. |

## II. <u>Summary</u>

| N | Field | Content |
|---|-------|---------|
| I.07 | **Warning in accordance with Article 6(7), second subparagraph, of Regulation (EU) 2023/1114** | The summary should be read as an introduction to the crypto-asset white paper.<br><br>The prospective holder should base any decision to purchase the asset-refenced token on the content of the crypto-asset white paper as a whole and not on the summary alone.<br><br>The offer to the public of the crypto-asset does not constitute an offer or solicitation to purchase financial instruments and any such offer or solicitation can be made only by means of a prospectus or other offer documents pursuant to the applicable national law. The crypto-asset white paper does not constitute a prospectus as referred to in Regulation (EU) 2017/1129 of the European Parliament and of the Council (36) or any other offer document pursuant to Union or national law. |
| I.08 | **Characteristics of the crypto-asset** | This document serves as an introduction to the Bitcoin network and the bitcoin crypto-asset, aimed at providing potential holders with essential information about Bitcoin.<br><br>Bitcoin is a peer-to-peer payments network offering a method to transact, store value and protect against debasement without reliance on a central governance or clearing authority. It is a digital ledger that is public and permissionless, which means that anyone can read the entire history of transactions in the ledger, and anyone can use the ledger to transact. This means that the network is transparent. However, it is also pseudonymic, that is, transactions and participants are identified by addresses rather than real-world identities. The network is not anonymous, but one real-world identity may correspond to multiple addresses and vice versa.<br><br>The unit of account in the Bitcoin network is "bitcoin", peer-to-peer electronic cash, each unit of which is subdivisible in 100 million units called "satoshis". A total of 21 million bitcoin will ever come into existence; this limited supply intends to act as a deflationary force over time, leading to long-term protection against debasement. Transactions in bitcoin are grouped in packets called "blocks" appended consecutively.<br><br>Transactions with bitcoin, once completed, are characterised by a strong sense of immutability and irreversibility. This is because editing the chain of blocks requires a very strong form of |
| | | "consensus" among network participants, and the roles and economic incentives in the protocol are designed in such a way that make double-spending attacks infeasible. A key component in this consensus is the "proof of work" algorithm, which attaches the likelihood of earning the right to submit the next block to the amount of hashing ("work", which, controlling for computational efficiency, equals energy) undertaken, and not to the number of digital identities issuing votes, which increases security. As hashing is an economically costly activity, and dishonest transaction validators' blocks will be rejected even if they have undertaken the work, the protocol places strong incentives for honest network maintenance without need for a central authority. |
| I.09 | **Key information about the offer to the public or admission to trading** | n/a |

**Part A: Information about the offeror or the person seeking admission to trading**

This section should be completed by the offeror or person seeking admission. As this whitepaper is written as an academic exercise, it is not applicable. We reproduce the [ESMA PoC](#) for guidance:

| N | Field | Content |
|---|---|---|
| A.01 | **Name** | Free alphanumeric text |
| A.02 | **Legal form** | ISO standard 20275 'Financial Services – Entity Legal Forms (ELF) |
| A.03 | **Registered address** | Free alphanumeric text |
| A.04 | **Head office** | Free alphanumeric text |
| A.05 | **Registration date** | YYYY-MM-DD |

| A.06 | **Legal entity identifier** | Legal entity identifier as defined in ISO 17442, e.g. 254900ARU0VC1WY6GJ71 |
|---|---|---|
| A.07 | **Another identifier required pursuant to applicable national law** | TBD |
| A.08 | **Contact telephone number** | 123456789 |
| A.09 | **E-mail address** | Free alphanumeric text |
| A.10 | **Response time** | Numerical {INTEGER-3} |
| A.11 | **Parent company** | Legal entity identifier as defined in ISO 17442 or another identifier required pursuant to applicable national law, e.g. 254900ARU0VC1WY6GJ71 |
| A.12 | **Members of management body** | Free alphanumeric text |
| A.13 | **Business activity** | Free alphanumeric text |
| A.14 | **Business activity of parent company** | Free alphanumeric text |
| A.15 | **Newly established** | 'true' – Yes<br>'false' – No |
| A.16 | **Recent financial condition** | Free alphanumeric text |
| A.17 | **Financial condition since registration** | Free alphanumeric text |

## Part B: Information about the issuer, if different from the offeror or person seeking admission to trading

This section should be completed by the issuer. As this whitepaper is written as an academic exercise, it is not applicable.

Moreover, Bitcoin does not have an issuer. Bitcoin units ("bitcoin") are issued to bitcoin validators ("miners") through a stochastic process that statistically rewards the ones that have spent the most effort ("work") in the consensus mechanism. Issuance of new bitcoin is open to anyone who wants to contribute to the security of the network via mining, and thus identities involved in issuance are not fixed.

We reproduce the [ESMA PoC](ESMA PoC) for guidance:

| N | Field | Content |
|---|---|---|
| B.01 | **Name** | Free alphanumeric text |
| B.02 | **Legal form** | Free alphanumeric text |
| B.03 | **Registered address** | Free alphanumeric text |
| B.04 | **Head office** | Free alphanumeric text |
| B.05 | **Registration date** | YYYY-MM-DD |
| B.06 | **Legal entity identifier** | Legal entity identifier as defined in ISO 17442, e.g. 254900ARU0VC1WY6GJ71 |
| B.07 | **Another identifier required pursuant to applicable national law** | TBD |
| B.08 | **Parent company** | Legal entity identifier as defined in ISO 17442 or another identifier required pursuant to applicable national law, e.g. 254900ARU0VC1WY6GJ71 |
| B.09 | **Members of management body** | Free alphanumeric text |

| B.10 | **Business activity** | Free alphanumeric text |
|---|---|---|
| B.11 | **Business activity of parent company** | Free alphanumeric text |

**Part C: Information about the operator of the trading platform in cases where it draws up the crypto-asset white paper**

This section should be completed by the offeror or person seeking admission. As this whitepaper is written as an academic exercise, it is not applicable. We reproduce the [ESMA PoC](#) for guidance:

| N | Field | Content |
|---|---|---|
| C.01 | **Name** | Free alphanumeric text |
| C.02 | **Legal form** | Free alphanumeric text |
| C.03 | **Registered address** | Free alphanumeric text |
| C.04 | **Head office** | Free alphanumeric text |
| C.05 | **Registration date** | YYYY-MM-DD |
| C.06 | **Legal entity identifier** | Legal entity identifier as defined in ISO 17442, e.g. 254900ARU0VC1WY6GJ71 |
| C.07 | **Another identifier required pursuant to applicable national law** | TBD |
| C.08 | **Parent company** | Legal entity identifier as defined in ISO 17442 or another identifier required pursuant to applicable national law, e.g. 254900ARU0VC1WY6GJ71 |

| C.09 | **Reason for crypto-asset white paper preparation** | Free alphanumeric text |
|---|---|---|
| C.10 | **Members of management body** | Free alphanumeric text |
| C.11 | **Operator business activity** | Free alphanumeric text |
| C.12 | **Business activity of parent company** | Free alphanumeric text |
| C.13 | **Other persons drawing up the crypto-asset white paper according to Article 6(1), second subparagraph, of Regulation (EU) 2023/1114** | Free alphanumeric text |
| C.14 | **Reason for drawing the white paper by persons referred to in Article 6(1), second subparagraph, of Regulation (EU) 2023/1114** | Free alphanumeric text |

## Part D: Information about the crypto-asset project

| N | Field | Content |
|---|---|---|
| D.1 | **Crypto-asset project name** | Bitcoin |
| D.2 | **Crypto-assets name** | bitcoin |

| D.3 | **Abbreviation** | BTC, XBT[3] |
|---|---|---|
| D.4 | **Crypto-asset project description** | Bitcoin is a purely peer-to-peer form of electronic cash, enabling online payments without reliance on any financial institution. As described in the summary in compliance with Section 6.7, the project prioritises transparency of transactions, security and immutability through a multi-step process of mining, validation and consensus, and permissionless read and write access to the protocol with nodes being able to rejoin the network at will. |
| D.5 | **Details of all natural or legal persons involved in the implementation of the crypto-asset project** | 1. **Creator**: Bitcoin was originally created by Satoshi Nakamoto, a person or group of people whose identity is not established and is considered to no longer be involved in the Bitcoin Core project.<br><br>2. **Advisors**: n/a. Bitcoin does not have formal advisors in the traditional sense of a crypto-asset project. Its development and strategic direction are influenced by the contributions and discussions within the Bitcoin community, including developers, miners, and users participating in forums like the BitcoinTalk forum and the bitcoin-dev mailing list.<br><br>3. **Development team**: Since Bitcoin is open-source, it doesn't have a formal, centralised development team. However, the Bitcoin Core software and implementation of the Bitcoin protocol is maintained and improved by a group of volunteer developers worldwide. There is a publicly available list of Bitcoin Core contributors in the [Github repository](#), including developers' usernames, with some choosing to provide their real names and others remaining anonymous. As of the date of publication of this whitepaper, this list includes 96 usernames. As this whitepaper is merely an educational instrument, we will omit listing their usernames, legal names and addresses. However, one should note that Bitcoin Improvement Proposals (BIPs) are only included in the Github repository after being submitted to the dev@lists.linuxfoundation.org mailing list, managed by The Linux Foundation. Its legal |

---

[3] Note: Bitcoin does not have a central governing authority, and participants may disagree about which versions of the protocol to support. These disagreements may result in the network splitting into two, in such a way that the two networks share a common chain of blocks and set of rules up until a point, and diverge in one or both dimensions thereafter. This event is known as a "fork" and each of the resulting networks is also denominated a "fork". Due to the lack of central authority, no fork is more legitimate than any other, and they can all be called "Bitcoin". For regulatory compliance purposes, this whitepaper focuses on the "Bitcoin Core" fork (BTC), but all Bitcoin forks are Bitcoin and emerge from Satoshi Nakamoto's original Bitcoin whitepaper. Tickers for other Bitcoin forks, not covered by this whitepaper include Bitcoin Cash (BCH), Bitcoin SV (BSV), Bitcoin Gold (BTG), Bitcoin Diamond (BCD), Bitcoin Scrypt (BTCS), BitcoinZ (BTCZ), Bitcoin Vault (BTCV), Super Bitcoin (SBTC), Bitcoin Private (BTCP), and Bitcoin Atom (BCA).

| | | |
|---|---|---|
| | | address is 548 Market St PMB 57274 San Francisco, California 94104-5401.<br><br>4. **Crypto-asset service providers**: n/a. There are no service providers operating formally with the Bitcoin project, because as a decentralised project there is no entity to formally engage with. There are entities such as exchanges (e.g., Coinbase, Binance), wallet providers (e.g., Trezor, Ledger), payment processors (BitPay, Coinbase Commerce), mining pools (F2Pool, Antpool), Layer 2 ecosystem building entities (Lightning Labs, Stacks Foundation) that independently choose to leverage the Bitcoin protocol and are, in a broad sense, relevant for its growth, but this whitepaper will not identify them as they are not formal providers to the project. Nonetheless, the Bitcoin Core repository is currently maintained in GitHub, a developer platform owned by GitHub, Inc., with a legal address at 88 Colin P Kelly Jr Street San Francisco, CA 94107 United States. |
| D.6 | **Utility Token Classification** | false |
| D.7 | **Key Features of Goods/Services for Utility Token Projects** | n/a |
| D.8 | **Plans for the token** | **Past milestones:**<br><br>- 2008: Publication of the Bitcoin whitepaper.<br>- 2009: Launch of the Bitcoin network with the mining of the genesis block (block number 0) on January 3rd.<br>- 2010: Bitcoin is used to make a real-world transaction for the first time when two pizzas are purchased for 10,000 BTC.<br>- 2012 (November): The first Bitcoin halving reduces the block reward from 50 BTC to 25 BTC.<br>- 2016 (July): The second Bitcoin halving reduces the block reward from 25 BTC to 12.5 BTC.<br>- 2017 (August): Implementation of Segregated Witness (SegWit).<br>- 2020 (May): The third Bitcoin halving reduces the block reward from 12.5 BTC to 6.25 BTC.<br>- 2021 (November): Activation of Taproot and introduction of Schnorr signatures. |
| | | **Future milestones:** <br><br> Unlike other crypto-asset projects, Bitcoin does not have a roadmap. Future milestones are not determined, and depend entirely on acceptance by the economic majority of Bitcoin users running full nodes. Currently, the only mechanism to submit proposals for changes to the protocol are Bitcoin Improvement Proposals (BIPs). BIPs are first discussed with, and filtered and copy-edited by the developer team described above, and then discussed openly by the community. A BIP becomes an accepted standard until its status becomes "Final" or "Active" and it becomes part of the protocol if it achieves consensus by the majority of users. <br><br> The repository of BIPs is located in [Github](). At the day of publication of this whitepaper, 389 BIPs are listed. These BIPs hold different status: active (2), final (24), draft (44), withdrawn (3), rejected (14), proposed (7), replaced (2), deferred (1), obsolete (0). <br><br> In general, outstanding BIPs are focused on enhancing functionality, security, and efficiency. Recent efforts emphasise the development of new features like Schnorr signatures and Taproot for improved privacy and scalability, alongside proposals for making Bitcoin more user-friendly through advancements in address formats and transaction signing processes. These proposals reflect ongoing efforts to refine and expand Bitcoin's capabilities to meet evolving user needs and technological standards. There are a number of proposals and areas of debate referring to the implementation of Layer-2 systems built on top of Bitcoin such as the Lightning Network and Stacks, and other innovations such as Ordinal Theory. Some may see these areas of work as a part of the "crypto-asset project" but others may not, as they are not a part of Bitcoin Core. |
| D.9 | **Resource Allocation** | **Resources already allocated to the project:** <br><br> 1. **Developer Time:** Volunteer developers contribute to the codebase. <br><br> 2. **Mining Infrastructure:** Significant investments by miners in computing hardware and electricity to secure the network and process transactions. At the time of publication of this whitepaper, the hash rate of the Bitcoin Network is approximately 614 exa-hashes per second (EH/s), which assuming an average of 100 tera-hashes per second (TH/s) per mining ASIC (application-specific |
| | | integrated circuit) roughly equals 6 million ASICs in hardware plus accompanying infrastructure.<br><br>3. **User nodes:** There are more than 10,000 Bitcoin nodes storing a copy of the entire blockchain and participating in consensus. Some sources place this figure closer to 100,000.<br><br>4. **Community Support:** Education, advocacy, and development support are provided by a global community, including non-profit organisations, research institutions, and individual contributors. |
| D.10 | **Planned Use of Collected Funds or Crypto-Assets** | n/a. No funds have been formally collected. Most of Bitcoin's initial development was self-funded by Satoshi Nakamoto, who may have received a donation between USD 2,000 and USD 3,600 from an anonymous donor to "pay for website hosting costs and other incidentals"[4]. |

### Part E: Information about the offer to the public of crypto-assets or their admission to trading

This section should be completed by the offeror or person seeking admission. As this whitepaper is written as an academic exercise, most fields are not applicable.

| N | Field | Content |
|---|---|---|
| E.1 | **Public Offering and/or Admission to trading** | n/a |
| E.2 | **Reasons for Public Offer and/or Admission to trading** | n/a |
| E.3 | **Fundraising Target** | n/a |

---

[4] Kuhn, D. (2024) "5 Things Satoshi Nakamoto Correctly Predicted About Bitcoin". CoinDesk. Available at https://www.coindesk.com/consensus-magazine/2024/02/26/5-things-satoshi-nakamoto-correctly-predicted-about-bitcoin/

| E.4 | **Minimum Subscription Goals** | n/a |
|---|---|---|
| E.5 | **Maximum Subscription Goal** | n/a |
| E.6 | **Oversubscription Acceptance** | n/a |
| E.7 | **Oversubscription Allocation** | n/a |
| E.8 | **Issue Price** | n/a |
| E.9 | **Official currency or any other crypto-assets determining the issue price** | n/a |
| E.10 | **Subscription fee** | n/a |
| E.11 | **Offer Price Determination Method** | n/a |
| E.12 | **Total Number of Offered/Traded CryptoAssets** | A total of 21 million bitcoin will ever be issued. This represents the upper limit of crypto-assets for trading. |
| E.13 | **Targeted Holders** | n/a |
| E.14 | **Holder restrictions** | n/a |
| E.15 | **Reimbursement Notice** | n/a |
| E.16 | **Refund Mechanism** | n/a |
| E.17 | **Refund Timeline** | n/a |
| E.18 | **Offer Phases** | n/a |
| E.19 | **Early Purchase Discount** | n/a |
| E.20 | **Time-limited offer** | n/a |

| E.21 | **Subscription period beginning** | n/a |
|---|---|---|
| E.22 | **Subscription period end** | n/a |
| E.23 | **Safeguarding Arrangements for Offered Funds /CryptoAssets** | n/a |
| E.24 | **Payment Methods for Crypto-Asset Purchase** | n/a |
| E.25 | **Value Transfer Methods for Reimbursement** | n/a |
| E.26 | **Public Offers** | n/a |
| E.27 | **Right of Withdrawal** | n/a |
| E.28 | **Transfer of Purchased Crypto-Assets** | n/a |
| E.29 | **Transfer Time Schedule** | n/a |
| E.30 | **Purchaser's Technical Requirements** | To hold bitcoin, a purchaser needs to directly manage a Bitcoin-compatible wallet ("self-custody") and its private keys, or have a third party manage such a wallet and keys. Bitcoin wallets may be "cold wallets", disconnected from the internet (paper wallets, sound wallets, hardware wallets) or "hot wallets", connected to the internet (software wallet apps or on-line browser wallets). |
| E.31 | **CASP name** | n/a |
| E.32 | **Placement form** | n/a |
| E.33 | **Trading Platforms** | n/a |
| E.34 | **Trading Platforms Access** | n/a |
| E.35 | **Involved costs** | n/a |

| E.36 | **Offer Expenses** | n/a |
|---|---|---|
| E.37 | **Conflicts of Interest** | n/a |
| E.38 | **Applicable law** | n/a |
| E.39 | **Competent court** | n/a |

**Part F: Information about the crypto-assets**

| N | Field | Content |
|---|---|---|
| F.1 | **Crypto-Asset Type** | Under the MiCA taxonomy, bitcoin is a crypto-asset of the "other" type. |
| F.2 | **Crypto-Asset Characteristics** | ● **ISO 24165 DTI code**: 4H95J0R2X<br>● **ISO 24165 FFG DTI**: V15WLZJMF<br>● **Free alphanumeric text**: The crypto-asset bitcoin is a "native digital token" using blockchain technology (public distributed ledger technology), with the hash algorithm Double SHA-256. Its genesis block hash 000000000019d6689c085ae165831e934ff763ae46a2a6c172b3f1b60a8ce26f, and genesis block UTC Timestamp 2009-01-03T18:15:05, long name "Bitcoin", short name "BTC" and "XBT", token reference URL https://github.com/bitcoin/bitcoin, unit multiplier 100,000,000, and the underlying asset external identifier ISIN: XTV15WLZJMF0. |
| F.3 | **Crypto-Asset Functionality Description** | Bitcoin are a means of payment (electronic cash) and a digital asset that may serve as a store of value, specifically protection against debasement. As a digital asset global in nature, it enables sending and receiving payments anywhere in the world at any time. It also transferred in a peer-to-peer manner, without need for financial intermediation. Because of this and the ability to self-custody the asset, it offers protection against confiscation. The crypto-asset does not offer other functionalities such as |

| | | accruing interest or staking rewards, awarding governance rights, entitling to airdrops. |
|---|---|---|
| F.4 | **Planned Application of Functionalities** | Functionalities described in section F.3 are already implemented since the crypto-asset's launch. For a discussion of other future improvements, see section D.8 "Future milestones". |

**Part G: Information on the rights and obligations attached to the crypto-assets**

| N | Field | Content |
|---|---|---|
| G.1 | **Purchaser Rights and Obligations** | Purchasers of bitcoin acquire property over the asset, which includes the bundle of rights under the applicable legal systems (right to use, right to transfer, etc.). Notably, bitcoin is dependent on the Bitcoin protocol rules – with any user being free to run the protocol fork implementation that they choose. The user thus holds the right to utilise the asset but does not hold a right to force others to run protocol implementations that they do not wish to run. Obligations are determined by the applicable legal system but there are no obligations intrinsic to the purchase of the asset itself. |
| G.2 | **Exercise of Rights and obligations** | Depends on the applicable legal system. |
| G.3 | **Conditions for modifications of rights and obligations** | There are no rules to create or modify rights and obligations for bitcoin purchasers, and the only conceivable path towards that would be a contractual arrangement among every single bitcoin owner. |
| G.4 | **Future Public Offers** | Bitcoin's issuance is predefined by its protocol, with a total cap of 21 million BTC, mined at a predictable rate. Currently, this rate is 6.25 BTC per block mined, with a block being mined approximately every 10 minutes (time is measured from timestamps, and difficulty is adjusted upwards or downwards to trend towards 10 minutes, with a difficulty adjustment every 2,016 blocks). This new issuance rate ("block subsidy") will be halved approximately in April 2024 (to 3.125 BTC per block) and will continue to be halved every 210,000 blocks (approximately four years) as it has in the past, until the last halvings. Three halvings have happened before the publication of |

|  |  | this whitepaper and another 29 are scheduled to occur. With the 32nd halving, the block reward will be reduced to 1 satoshi, and with the 33nd halving, to 0.5 satoshis. However, as the Bitcoin protocol does not support unit subdivisions below 1 satoshi, the last halving will entail the finalisation of the issuance of bitcoin. There is no issuer in the traditional sense; thus, future offers and retention do not apply as they would in conventional securities or crypto-assets issued by a central entity. |
|---|---|---|
| G.5 | **Issuer Retained Crypto-Assets** | Unlike other crypto-asset projects, Bitcoin did not have a "pre-mine" and hence no assets were retained by any "issuer". |
| G.6 | **Utility Token Classification** | false |
| G.7 | **Key Features of Goods/ Services of Utility Tokens** | n/a |
| G.8 | **Utility Tokens Redemption** | n/a |
| G.9 | **Non-Trading request** | false |
| G.10 | **Crypto-Assets purchase or sale modalities** | There are no restrictions to the purchase or sale of bitcoin other than those imposed by regulation. |
| G.11 | **Crypto-Assets Transfer Restrictions** | There are no technical restrictions to bitcoin transferability other than the avoidance of double-spending and following the protocol rules. Legally, regulation may restrict the transferability of bitcoin. |
| G.12 | **Supply Adjustment Protocols** | In Bitcoin, there are no protocols to vary supply in response to changes in demand for the crypto-asset. If bitcoin demand increases, bitcoin issuance does not increase, and if bitcoin demand decreases, bitcoin issuance does not decrease in response. Rather, bitcoin issuance is determined algorithmically, according to the "block subsidy" rules. Block subsidy was originally 50 bitcoin and is scheduled to cut in half every 210,000 (approximately every four years). Currently, the subsidy is 6.25 bitcoin. Approximately in the year 2140, the last bitcoin will be |
| | | mined, with cumulative supply totalling 21 million. This corresponds to 63 halvings. Once halving number 32 is reached however, bitcoin issuance will be so small that it will overcome the 8 decimal limit and will not be describable with Bitcoin's current parameters. Mathematically, the formula describing bitcoin issuance can be outlined as follows: $$\sum_{i=0}^{32} 210,000 \left(\frac{50}{2^i}\right)$$ In this formula, 32 is the number of halvings that will occur before the last satoshi is mined, i=0 represents the starting index for the reward era (each 210,000 block period is a "halving era"), 210,000 is the number of blocks between halvings, and 50 is the number of bitcoin issued per block during starting era i=0. |
| G.13 | **Supply Adjustment Mechanisms** | Although there are no direct mechanisms to adjust to changes in the demand for the crypto-asset, there is a mechanism to adjust to changes in mining activity. As mining constitutes one of the channels through which purchasers may acquire bitcoin, mining activity may indirectly and partially reflect the demand for the underlying bitcoin. New bitcoin are issued every time a miner finds a valid nonce (see section H.4 for a description of the process). Finding a nonce is a stochastic trial-and-error process depending on the amount of hashing undertaken relative to the difficulty level. If the demand for bitcoin increases and so does the global bitcoin hash rate, valid nonces will be found with a frequency shorter than 10 minutes, leading to an increased supply. The Bitcoin protocol is designed to compensate for this and ensure that a new block is produced every 10 minutes on average. This adjustment is produced once every 2,016 blocks (approximately every 2 weeks). As every time that a new block is added to the blockchain, it includes a timestamp reflecting when it was mined, the protocol can calculate the time it took to mine 2,016 blocks and compare it with a "target time" of 20,160 minutes (2 weeks). If the actual time is below the target time, difficulty is adjusted upwards; if it is above the target time, downwards. The difficulty adjustment formula can be described as follows: $$New\ difficulty\ =\ Old\ difficulty\ \times \frac{Actual\ time\ of\ last\ 2,016\ blocks}{Target\ time\ of\ 2,016\ blocks}$$ This ensures that the issuance schedule described in section G.12 is respected. For a more detailed description of the mining process, |

| | | see section H.4. |
|---|---|---|
| G.14 | **Token Value Protection Schemes** | n/a |
| G.15 | **Compensation schemes** | n/a |
| G.16 | **Applicable law** | Due to Bitcoin's decentralised nature, it is not governed by the laws of any single jurisdiction, making its legal framework complex and varied globally. The applicable law for matters related to Bitcoin transactions would depend on the jurisdiction of the involved parties. |
| G.17 | **Competent court** | The competent court for matters related to Bitcoin transactions would depend on the jurisdiction of the involved parties. |

**Part H: Information on the underlying technology**

| N | Field | Content |
|---|---|---|
| H.1 | **Distributed ledger technology** | Public permissionless blockchain, relying on "Nakamoto Consensus" which leverages the "proof of work" algorithm relying on stochastic selection of block-proposers by turns (see section H.4). The blockchain is maintained by:<br><br>● Full nodes: They store the entire history of blockchain transactions, validate blocks and transactions according to consensus rules, and relay this information to other nodes. A subset of full nodes are also mining nodes.<br>    ○ Mining nodes: They conduct the "work" described in section H.4.<br>● Lightweight or Simplified Payment Verification (SPV) nodes: They do not store the entire blockchain but can verify transactions by downloading only the headers of blocks.<br>● Listening nodes: They pass new transactions and blocks to other nodes, facilitating efficient data propagation across the |
| | | network. They help in maintaining the network's connectivity and resilience. |
| H.2 | **Protocols and technical standards** | - Hashing algorithm: SHA-256.<br>- Wallet key generation and transaction signature algorithm: ECDSA (Elliptic Curve Digital Signature Algorithm).<br>- Node communication protocol: peer-to-peer (P2P).<br>- Scripting language for transactions: Bitcoin Script (stack-based).<br>- Primary development language: C++.<br>- Block size: 1 megabyte (MB).<br>- Block weight: 4 million weight units (4 MB).<br>- Block time: 10 minutes. |
| H.3 | **Technology Used** | n/a. Bitcoin is not an asset-referenced token. See section E.30 for a description of wallets. |
| H.4 | **Consensus Mechanism** | The Bitcoin protocol relies on "Nakamoto consensus", a consensus process involving the "Proof of Work" (PoW) algorithm. The Nakamoto consensus begins with miners selecting pending transactions from a pool, bundling them into a new block with a nonce. They compute the block's hash to match the network's difficulty target, adjusting to maintain a near 10-minute block interval. This process involves nonce iterations to solve the challenge posed by the current difficulty level.<br><br>This process works as follows: Miners have packeted transactions in a block. The block contains a block header (as well as signed transaction data and a transaction counter) which includes the version number, the previous block's hash, the Merkle root of all transactions in the block, the timestamp, the difficulty target, and the "nonce", The nonce (number only used once) is an input variable that miners adjust in their attempt to find a block hash that meets the network's difficulty target. Essentially, the nonce is a number that miners change with each hashing attempt, allowing them to generate different hashes from the same block data until they discover a hash below the difficulty threshold, successfully mining a new block. Miners adjust the nonce in the block header and repeatedly hash the header until they find a hash that meets the difficulty target. This process is essentially trial and error, with the nonce serving as the variable |

component allowing for different hash outputs until the correct one is found.

Through this process, miners solve for a hash that is less than or equal to the network's current difficulty target (mining). An acceptable hash is one that starts with a certain number of zeros, the exact count of which is dictated by the difficulty level. The difficulty adjusts to ensure the average time between blocks remains around 10 minutes, aiming for a balance between network security and processing efficiency.

Once a valid nonce yielding an acceptable hash is discovered, the miner has the "proof of work" and broadcasts the block to the network. Other nodes independently verify the block's transactions and hash, ensuring it adheres to protocol rules (and that the miner has done the work), including transaction signature validation and hash target compliance. Upon verification, nodes append the block to their blockchain version.

Multiple chains can emerge when two miners solve the "puzzle" nearly simultaneously, leading to two valid blocks. Nodes may receive these blocks at different times due to network latency delaying block propagation, creating temporary divergences or "forks" in the blockchain as miners work on top of different blocks temporarily. Forks in Bitcoin can also occur due to differences in node software updates, leading to incompatibilities in block validation rules. Additionally, network latency can delay block propagation, causing nodes to work on different blocks temporarily. Users or miners rejecting a block due to perceived flaws or deviations from the consensus rules can further contribute to the creation of forks. This selective acceptance can lead to temporary forks as different parts of the network wait for a block they agree upon, highlighting the role of network consensus in determining the blockchain's state.

The network resolves this by extending the chain that receives the next block first, as miners generally work on the longest available chain. Over time, the chain with the most accumulated work, typically the one extended first (although it can also be another one if the majority of the users choose to reject certain blocks for other reasons), becomes accepted by the network as the definitive transaction history (and the other chain is abandoned, or "orphaned"). This process ensures that despite temporary forks, a single, agreed-upon chain prevails, reinforcing the security and decentralisation of Bitcoin.

Finality in Bitcoin, or the assurance that transactions cannot be altered, is probabilistically achieved as more blocks are added atop a transaction. Generally, six confirmations (approximately one hour) are

| | | | |
|---|---|---|---|
| | | | considered sufficient to attain a high degree of finality, significantly reducing the risk of transaction reversal due to chain reorganisations. |
| | | | This mechanism secures the network against double-spending and confirms transaction authenticity without central oversight, as nodes recognise the most worked-on chain as the accurate transaction history, establishing trust and consistency in the decentralised network. The whole process of mining, broadcasting, independent verification, decentralised selection which block to build on, and the longest-chain rule is called Nakamoto consensus, with proof of work being one of its key components during the mining and verification steps. |
| H.5 | | **Incentive Mechanisms and Applicable Fees** | The incentive to secure transactions refers to the incentives for miners and for user nodes. Miners are incentivised to mine, a very costly activity, via "block rewards": a reward assigned to the miner that finds the right nonce (and that is honest and lucky enough to be working in the longest chain). |
| | | | Block rewards have two components, the block subsidy and transaction fees. Transaction fees are not predetermined. Users that wish to send a bitcoin transaction to the transaction pool decide by themselves the size of the payment that the miner who selects their transaction (and is successful to find the right nonce afterwards) will receive – and are even free to set the transaction fee to zero. However, miners are also free to select the transactions that they prefer, and will typically select higher paying transactions first. Because block space is scarce, this creates a market for block space, where the price of block space, i.e. transaction fees, is determined by supply and demand. |
| | | | The Bitcoin network needs to achieve "critical mass" for transaction fees to be sufficient to incentivise sufficient mining activity to keep the network functional and secure. Hence, there are network effects at play: demand for block space depends on demand for Bitcoin usage; however, demand for Bitcoin usage depends in turn on network security (which depends on mining activity) as well as on the passage of time. To avoid a negative feedback loop, kickstart the network and enable it to achieve the necessary network effects and maturity to be self-sustaining, the block subsidy is implemented as a temporary measure (and the only mechanism to bring new bitcoin into existence). However, as the network matures, the subsidy becomes less necessary. For this reason, the protocol includes the decreasing subsidy schedule marked by the halvings described above, which also acts as a supply limitation mechanism that allows bitcoin to offer protection against debasement. Hence the incentives seek to incentivise early mining participation while controlling inflation. |

|  |  | This incentive system is underpinned by game theory, ensuring miners act in the network's best interest. The costly proof of work required to mine blocks means that any attempt to undermine the network, such as by validating fraudulent transactions, would result in substantial losses for a miner in terms of wasted computational resources and foregone legitimate rewards. As a result, miners are economically motivated to maintain the network's integrity, with the consensus mechanism ensuring that only the longest chain of blocks, representing the majority computational effort, is accepted as true. This aligns individual miners' incentives with the collective goal of securing the Bitcoin network.

User nodes, or full nodes, perform transaction verification out of a commitment to maintain the network's integrity and security. While they don't receive direct financial incentives like miners, their role upholds the decentralised ethos of Bitcoin, ensuring transactions are valid and the blockchain remains accurate and resistant to attacks. Their incentive may be ideological, to support a secure, decentralised currency system, and/or practical, as it ensures their own transactions and those they rely on are accurately processed and recorded. |
|---|---|---|
| H.6 | **Use of Distributed Ledger Technology** | false[5] |
| H.7 | **DLT Functionality Description** | n/a |
| H.8 | **Audit** | false[6] |
| H.9 | **Audit outcome** | n/a |

---

[5] There is no single third party operating the issuance, transfer, or storage of bitcoin. The entire set of bitcoin miners govern its issuance, the entire set of users govern its storage, and the entire set of miners and users govern its transfer.

[6] Bitcoin code has not undergone a formal audit, but this is not applicable as the code is public, open-source and is constantly reviewed by the community.

**Part I: Information on the risks**

| N | Field | Content |
|---|---|---|
| I.1 | **Offer-Related Risks** | Specific offerors or persons seeking admission to trading should elaborate on risks specific to the circumstances, but risks are more reduced than in other crypto-asset projects because there is no central entity raising capital or making governance decisions, such that their financial sustainability or governance practices may have an impact on the public. As bitcoin is not an asset-referenced token, there is no reserve of assets, financial risks or liabilities to be managed. More generally however, Bitcoin network decentralisation and immutability, while offering many advantages, also presents risks related to usability and consumer protection. Without a central authority with the technical ability to reverse transactions, payments made under duress or deceit, in error, or in violation of the law, cannot be un-done. A court may order a new transaction to be made to refund or compensate an aggravated party, but the court does not have the technical ability to execute its decision without access to the private keys of the address at issue – a situation similar to physical cash. These risks can be managed by relying on intermediaries, especially regulated ones and with a capable customer support service. However, in this case the purchaser is exposed to the traditional risks of relying on third parties. |
| I.2 | **Issuer/ Offeror/ Admission Entity Differentiation** | true |
| I.3 | **Issuer-Related Risks** | Anyone can mine bitcoin, and hence there are no restrictions on who can issue it, as long as they follow protocol rules and provide the proof of work. This means that sanctioned parties whose income authorities seek to impair may not be prevented from engaging in this activity if they have the hardware and electricity necessary to mine. The mining industry has been professionalised over the past years and mining companies today are, usually, increasingly large, regulated and listed – however "rogue states" that choose to mine bitcoin are able to do so, and the bitcoin that they mine is fungible with bitcoin mined by others. |
| I.4 | **Crypto-Assets-related Risks** | 1. **Lack of underlying asset**: Not being asset-referenced and not having a cash flow to discount, the intrinsic value of a bitcoin is hard to determine. While it is a protection against debasement in the long term (assuming sustained demand |

|  |  | for the asset), in the short term this can entail price volatility for bitcoin.<br><br>2. **Irreversibility of transactions:** As explained before, the immutability of the ledger has many strengths but also introduces some risks related to payments in error, fraud, etc.<br><br>3. **Privacy:** Transaction details are publicly recorded on the blockchain, potentially exposing user activities. Although bitcoin are fungible and addresses are pseudonymic, the transparency and immutability of the entire history of bitcoin transactions enables forensics and intelligence to be developed that may allow tracing addresses to real world identities.<br><br>4. **Loss of private keys**: This represents a significant risk, as it results in the irreversible loss of access to Bitcoin holdings, underscoring the importance of secure key management practices.<br><br>5. **Monetary sovereignty:** By offering a protection against debasement and confiscation, Bitcoin also presents a risk to monetary sovereignty by the state, as individuals may choose to transact on Bitcoin instead of state-issued currency.<br><br>6. **Market Liquidity and Price Volatility:** Bitcoin markets may experience liquidity constraints and price volatility, leading to potential losses or market inefficiencies for investors and traders. |
|---|---|---|
| I.5 | **Project Implementation -Related Risks** | The decentralised nature of Bitcoin presents challenges for network development and upgrades. Achieving consensus for changes is complex, as it requires agreement across a diverse group of participants. This can lead to potential forks, where the blockchain diverges into separate paths, reflecting differing visions within the community, possibly impacting the crypto-asset's value. In addition, reluctance to change by the community may lead to excessive rigidities and adaptation problems to new market conditions or technical developments, and conversely an excessive tendency to modify protocol rules may affect the perceived immutability of the protocol and hence of the underlying crypto-asset's value. |
| I.6 | **Technology- Related Risks** | 1. **Protocol Vulnerabilities:** The Bitcoin protocol may be vulnerable to technical flaws or bugs, potentially leading to network disruptions, security breaches, or manipulation. |

2. **Cybersecurity Threats:** The Bitcoin network itself, Bitcoin wallets, exchanges, and other infrastructure are susceptible to cyberattacks, hacking attempts, and phishing scams, posing risks of theft, fraud, or loss of funds.

3. **Environmental Risk:** Bitcoin mining consumes vast amounts of energy, also from non-renewable sources, contributing to carbon emissions and exacerbating climate change. Additionally, e-waste generated from obsolete mining hardware may pose an environmental threat.

4. **Scalability Challenges**: The scalability of the Bitcoin network may be limited by its consensus mechanism and block size, leading to congestion, delays, and higher transaction fees during periods of high demand and consequently the risk of being leapfrogged by faster or cheaper architectures.

5. **Decreasing Security Budget Risk:** The security of the network highly depends on the rewards paid out to miners in the network. The rewards depend on transaction fees and block subsidy, whereas in the past the block subsidy has been the majority of miners' earnings. As roughly every four years the block subsidy halves, miners' earnings could shrink and increase risks of attacks.

6. **Regulatory and Compliance Risks:** Regulatory uncertainty and evolving legal frameworks may pose challenges for Bitcoin adoption and acceptance, leading to compliance risks for businesses and individuals using or investing in Bitcoin. This is particularly driven by the high energy requirements by the Bitcoin network which has led to regulatory considerations in the past.

| I.7 | **Mitigation measures** | |
|---|---|---|
| | | 1. **Protocol Vulnerabilities:** Regular code reviews, security audits, and protocol upgrades can help identify and address vulnerabilities. Engaging with the open-source community for peer review and collaboration can enhance the resilience of the Bitcoin network. |
| | | 2. **Cybersecurity Threats:** The lack of a central point of failure mitigates the possibility of cyberattacks on the Bitcoin network itself. Implementing robust cybersecurity measures, including encryption, multi-factor authentication, and cold storage solutions, can mitigate the risk of unauthorised access and protect user assets. Education and awareness programs can also help users recognize and avoid common security threats. |
| | | 3. **Environmental Risk:** The Bitcoin mining community should transition to renewable energy sources for mining operations, implement proper carbon accounting practices, and invest in responsible disposal and recycling of e-waste. Collaboration between stakeholders and technological innovation towards less environmentally harmful alternatives, e.g. to replacing hazardous waste in hardware can also help to reduce the environmental impact of Bitcoin. |
| | | 4. **Scalability Challenges:** Research and development efforts focused on scalability solutions, such as the Lightning Network or off-chain scaling solutions, aim to enhance the throughput and efficiency of Bitcoin transactions. Investing in infrastructure upgrades and network optimizations can improve scalability and user experience over time. Ongoing innovation by monitoring market trends and updating functionality accordingly should be ensured. |
| | | 5. **Decreasing Security Budget Risk:** An increase in transaction fees can render the network more resilient against attacks and help miners to continue their businesses. Potential other mitigation strategies such as issuing new coins or re-issuing old coins are discussed controversially in the community. |
| | | 6. **Regulatory and Compliance Risks:** Engaging with regulators, policymakers, and industry associations to advocate for clear and favourable regulatory frameworks can help reduce uncertainty and promote mainstream adoption of Bitcoin. Implementing robust compliance programs, including AML and KYC procedures, can mitigate regulatory risks and |
| I.7 | **Mitigation measures** | |

| | | enhance trust and transparency in Bitcoin transactions. Enhancing transparency on the energy consumption and GHG emissions of mining companies in accordance with commonly accepted carbon accounting frameworks such as the GHG Protocol. |
|---|---|---|

### Part J: Information on the sustainability indicators in relation to adverse impact on the climate and other environment-related adverse impacts

**Mandatory information on principal adverse impacts on the climate and other environment-related adverse impacts of the consensus mechanism**

**General information and key indicators**

CCRI - Crypto Carbon Ratings Institute incorporated as CCRI GmbH in Dingolfing (Germany), acting as sustainability data provider since no issuer or offeror can be identified for Bitcoin, is providing information on principal adverse impacts on the climate and other environment-related adverse impacts of the consensus mechanism used to validate transactions in Bitcoin (BTC) and to maintain the integrity of the distributed ledger of transactions.

The information covers the period from 01.01.2023 to 31.12.2023.

The validation of transactions in Bitcoin and the maintenance of the integrity of the distributed ledger of transactions has led to a total energy consumption of 121,134,042.6 kWh during 2023.

The validation of one transaction in Bitcoin has led to a total energy consumption of 44.47 kWh on average during 2023.

The validation of transactions in Bitcoin and the maintenance of the integrity of the distributed ledger of transactions has resulted in 61,311,929.06 tonnes GHG emissions, calculated based on sources owned or controlled by the DLT network nodes (scope 1), and indirect emissions from energy purchased by the DLT network nodes (scope 2), during 2023.

**Features of the consensus mechanism relevant for principal adverse impacts on the climate and other environment-related adverse impacts**

See Part H: Information on the underlying technology

| Climate and other environment-related indicators | | | | |
|---|---|---|---|---|
| 1 | 2 | 3 | 4 | 5 |
| | Adverse sustainability indicator | Metric | Source of information, review by third parties, use of data providers or external experts | Methodology to calculate metrics from information and data obtained |
| Energy | Energy consumption | 121,134,042.6 kWh | Provided by CCRI as external party | See CCRI API documentation |
| | Non-renewable energy consumption | 72.76%[7] | Provided by CCRI as external party | See CCRI API documentation |
| | Energy intensity | 44.47 kWh | Provided by CCRI as external party | See CCRI API documentation |
| GHG emissions | Scope 1 – Controlled | - | Provided by CCRI as external party | See CCRI API documentation |
| | Scope 2 – Purchased | 61,311,929.06 tCO2e[8] | Provided by CCRI as external party | See CCRI API documentation |
| | GHG intensity | 21.92 kg/Tx | Provided by CCRI as external party | See CCRI API documentation |

---

[7] For reference, the non-renewable energy share of the US is at 80.20% and at 72.21% worldwide according to IRENA (2023), Renewable Capacity Statistics 2023; & IRENA (2023), Renewable Energy Statistics 2023, The International Renewable Energy Agency, Abu Dhabi. The share of the non-renewable energy consumption is calculated based on mining location data provided by the Cambridge Center of Alternative Finance (CCAF). The data has been last updated by CCAF in January 2022. As soon as more up-to-date data becomes available, all affected indicators should be updated.

[8] The scope 2 emissions have been calculated solely using country or state grid emission factors in accordance with the location-based carbon accounting approach. According to the GHG Protocol, a market-based accounting approach can be conducted in addition if renewable energy claims can be verified. As soon as verifiable renewable energy claims become available on a large scale, scope 2 emission for the Bitcoin network can be added for comparison.

| Waste production | Generation of waste electrical and electronic equipment (WEEE) | 10,350 t | Provided by CCRI as external party | [See CCRI API documentation](#) |
|---|---|---|---|---|
| | Non-recycled WEEE ratio | 84.75% | Provided by CCRI as external party | [See CCRI API documentation](#) |
| | Generation of hazardous waste | 122 t | Provided by CCRI as external party | [See CCRI API documentation](#) |
| Natural resources | Impact of the use of equipment on natural resources | The impact on natural resources, such as water, is largely driven by the magnitude of the electricity consumed by the network and the related equipment. The annual water footprint is estimated at 1,909 gigalitres. | Provided by CCRI as external party | [See CCRI API documentation](#) |